\documentclass[aps,prl,showpacs,twocolumn,groupedaddress]{revtex4}  
\usepackage[latin1]{inputenc} 
\usepackage{graphicx,epsfig,epstopdf}

\begin{document} 
 
\title{Quantum Magnets under Pressure: Controlling Elementary Excitations 
in TlCuCl$_3$} 

\author{  
Ch. R\"uegg,$^{1}$
B. Normand,$^{2,3}$
M. Matsumoto,$^{4}$
A. Furrer,$^{5}$
D. McMorrow,$^{1}$
K. Kr\"amer,$^{6}$
H.--U. G\"udel,$^{6}$
S. Gvasaliya,$^{5}$
H. Mutka,$^{7}$
and M. Boehm$^{7}$} 
 
\affiliation{
$^1$ London Centre for Nanotechnology; University College London; London 
WC1E 6BT; UK \\ 
$^{2}$ D\'epartement de Physique; Universit\'e de Fribourg; CH--1700 
Fribourg; Switzerland \\
$^{3}$ Theoretische Physik; ETH--H\"onggerberg, CH--8093 Z\"urich; 
Switzerland \\
$^{4}$ Department of Physics; Faculty of Science; Shizuoka University; 
Shizuoka 422--8529; Japan \\
$^{5}$ Laboratory for Neutron Scattering; ETH Zurich and Paul Scherrer 
Institute; CH--5232 Villigen PSI; Switzerland \\
$^{6}$ Department of Chemistry and Biochemistry; University of Bern; 
CH--3000 Bern 9; Switzerland \\
$^{7}$ Institut Laue Langevin; BP 156; 38042 Grenoble Cedex 9; France}

\date{\today} 

\begin{abstract} 
We follow the evolution of the elementary excitations of the quantum 
antiferromagnet TlCuCl$_3$ through the pressure--induced quantum critical 
point, which separates a dimer--based quantum disordered phase from a 
phase of long--ranged magnetic order. We demonstrate by neutron 
spectroscopy the continuous emergence in the weakly ordered state 
of a low--lying but massive excitation corresponding to longitudinal 
fluctuations of the magnetic moment. This mode is not present in a 
classical description of ordered magnets, but is a direct consequence 
of the quantum critical point.
\end{abstract} 
 
 
\maketitle 

Although quantum fluctuations of both spin and charge degrees of 
freedom are the key to the essential physics of many challenging 
problems in condensed matter systems, the microscopic control of 
zero--point fluctuations has to date remained largely a theoretical 
abstraction. However, full control over the interaction parameters can 
now be effected in cold atomic condensates through the standing--wave 
amplitudes of the optical lattice. Similarly, in quantum magnets the 
exchange interactions can be controlled by the application of pressure, 
altering the effect of spin fluctuations. We follow this approach to 
investigate the physics of a quantum system whose fluctuations are 
``tuned'' in a continuous way.

\begin{figure}[t] 
\includegraphics[width=0.45\textwidth]{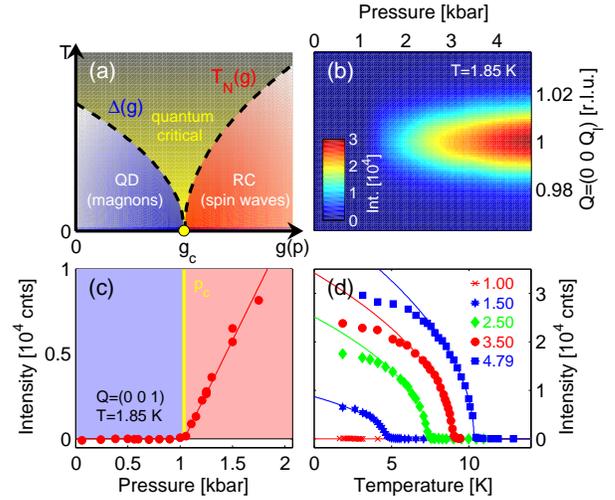}
\caption{(a) Generic phase diagram for a QPT occurring as a function 
of parameter $g(p) \propto \sum_j J_{ij}(p)/J(p)$. For a magnetic QPT, 
the characteristic energy scales in the QD and RC phases are respectively 
the spin gap $\Delta$ and N\'eel temperature $T_{\rm N}$, both of which 
vanish at the QPT. The nature of the lowest--lying excitations is given 
in parenthesis. (b--d) Pressure-- and temperature--dependence of the 
magnetic Bragg peak intensity at ${\mbox{\boldmath$Q$\unboldmath}}$
 = (0 0 1) in TlCuCl$_3$, which is proportional to the square of the 
order parameter $m_{s}$.} 
\label{fig1}  
\end{figure} 

The most dramatic manifestation of such control is the driving of a 
quantum phase transition [QPT, Fig.~1(a)] between two different ground 
states \cite{rchn}. Structurally dimerized $S = 1/2$ spin systems 
offer a particularly clean realization both of the magnetic field--induced 
QPT, which has been studied extensively in a number of materials 
\cite{Giamarchi08}, and of the qualitatively different magnetic QPT 
driven by hydrostatic pressure \cite{Sachdev08}. The Hamiltonian 
\begin{equation}
{\cal H}_{\rm H} = \sum_i J (p) {\bf S}_{i,l} {\bf \cdot S}_{i,r}
+ \sum_{ij;m,m'=l,r} J_{ij} (p) {\bf S}_{i,m} {\bf \cdot S}_{j,m'}
\end{equation}
contains pressure--dependent Heisenberg interactions $J(p)$ for intra-- 
and $J_{ij}(p)$ for interdimer bonds between spins ($l$,$r$) on dimers 
$i$ and $j$. Schematically, stabilization of the quantum disordered (QD) 
phase is driven by spin--singlet formation on the dimer units, while a 
weakening of this tendency leaves a renormalized classical (RC) phase, 
whose conventional properties are partially suppressed by dimer 
fluctuations \cite{Matsumoto04}. Because of its extremely low critical 
pressure, $p_c =$ 1.07 kbar \cite{Rueegg04}, the pressure--induced 
QPT in TlCuCl$_3$ \cite{Tanaka03} offers a unique opportunity to study 
static and dynamic properties throughout the quantum critical regime. 
Here we determine by high--resolution inelastic neutron scattering 
(INS) the spin excitation energies, spectral weights, lifetimes, and 
polarizations. 

The QD ground state is a spin singlet with three triplet excitation 
branches, which are fully gapped and, for a system with unbroken spin 
symmetry, degenerate. On the RC side but far from the QPT, one has 
effectively an ordered state with a rigid magnetic moment, whose 
key excitations are considered to be only two massless spin waves, 
the Goldstone modes corresponding to transverse (phase) fluctuations 
\cite{Matsumoto04,Rueegg04}. However, the RC phase takes on an 
increasingly exotic nature close to the continuous QPT, where the 
hallmarks of "classical" behavior are present only as a thin veneer 
superposed on the dimer--singlet background. We will show that a 
further low--lying excitation is present in this regime: the missing 
longitudinal mode \cite{ra,rnr} emerges as the small ordered 
moment becomes increasingly ''malleable'' on approaching the QPT 
systematically by controlling the applied pressure. Triplet modes in 
TlCuCl$_3$ have to date been measured only with rather coarse resolution, 
in the QD state \cite{Cavadini01} and deep within the RC phase at 7.3 kbar 
($p \gg p_c$) \cite{Rueegg04}, where the spin structure has also been 
elucidated \cite{Oosawa04}. Our data now connect these limits, providing 
unprecedented high--resolution information on the evolution of the 
excitation spectra across the QPT. 

High--quality single crystals of TlCuCl$_3$, with sample mass 1.5 g and 
a pressure--independent mosaic spread of 0.5$^\circ$, are grown by the 
Bridgman method. The INS experiments were performed on the triple--axis
spectrometers TASP (SINQ) and IN14 (ILL), working in constant 
final--energy mode with a focusing pyrolytic graphite analyzer/monochromator
and respective horizontal collimations 50'--open--open--open and 
open--60'--open--open. The instruments were operated respectively at $E_f
 =$ 3.5 meV and 3.0 meV, both giving elastic energy resolutions of 0.1 meV 
(full width at half maximum height) across the spectral range, and very 
clean background conditions. A cooled Be filter is positioned between 
the analyzer and the sample, which was housed in a He--gas pressure 
cell and in a standard cryostat operating at $T \ge 1.5$ K. INS 
measurements detect the components of magnetic fluctuations in the plane 
perpendicular to the momentum transfer ${\mbox{\boldmath$Q$\unboldmath}}$. 
Specifically, points ${\mbox{\boldmath$Q$\unboldmath}}$ = (0~0~1) and 
(0~4~0) give access to spin fluctuations in all three spatial directions, 
and thus for an ordered magnet contain both transverse and longitudinal 
excitations. The extent to which the individual mode polarizations are 
present at these wave vectors is determined by the magnetic structure of 
TlCuCl$_3$ above $p_c$, and our assignment (below) is fully consistent 
with the results reported in Ref.~\cite{Oosawa04}.

\begin{figure}[t] 
\begin{center}
\includegraphics[width=0.40\textwidth]{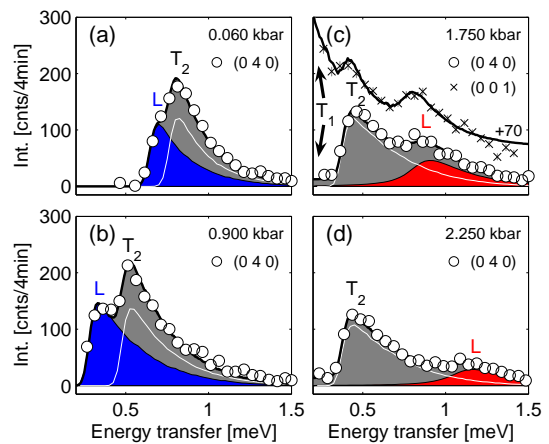}
\end{center}
\caption{INS spectra showing triplet excitations at $T = 1.85$ K and 
${\mbox{\boldmath$Q$\unboldmath}}$ = (0 4 0) for four different 
pressures across the QPT. Complementary data taken at 
${\mbox{\boldmath$Q$\unboldmath}}$ = (0 0 1) is shown in (c). }
\label{fig2}  
\end{figure}

We touch only briefly on static properties: Figs.~1(b--d) show the 
complete pressure-- and temperature--dependence of the ordered moment, 
$m_s$, measured through the magnetic Bragg peak intensity. A clear and 
continuous onset is visible in Fig.~1(b), showing that the QPT is of 
second order, with a linear pressure--dependence [Fig.~1(c)] on the 
RC side. The temperature--dependence shown in Fig.~1(d) is used to 
extract $T_{\rm N} (p)$ \cite{Rueegg04}.

Typical INS spectra for the spin excitations in TlCuCl$_3$ are shown 
in Fig.~2 for a number of pressure values. The INS intensities at the 
band minimum for $p < p_c$ [Figs.~2(a,b)] show a closing of the gaps: 
there are two resolved excitations, labeled T$_2$ and L. The intensities 
change dramatically on crossing the QPT [Fig.~2(c)], with T$_2$ 
essentially unchanged, an apparently massless mode T$_1$ becoming 
visible, and L opening a significant gap. At high pressures [Fig.~2(d)], 
the low--lying modes (T$_1$ and T$_2$) have unchanged energies, whereas 
the emerging gapped mode L has moved still higher, losing intensity and 
broadening simultaneously. Figure 3 summarizes the INS data for the 
properties of the excitations across the QPT, displaying clearly three 
key features: i) there is a low--pressure splitting of the triplet 
manifold; ii) the QPT is a second--order transition to within
experimental resolution in all quantities measured; iii) the two spin 
waves of the ordered phase, one of which is massive, are accompanied 
by a well--defined longitudinal mode, whose properties change continuously 
with applied pressure.

Figure 4 presents a complete characterization of the longitudinal mode, 
beginning with the intensity data [Fig.~4(a)] obtained from the red peaks 
shown in Figs.~2(c--d), from which the mode energy [Fig.~4(b)], integrated 
intensity [Fig.~4(c)], and full width at half maximum height [Fig.~4(d)] 
are extracted. On moving away from the QPT, the mode mass rises 
monotonically [Fig.~4(b)] and its intensity weakens [Fig.~4(c)], 
indicating a larger, stiffer magnetic moment [Fig.~1(c)], and hence a 
reduced effect of quantum fluctuations. The broadening is always small 
compared to the mass, and vanishes systematically on approaching the 
QPT [Fig.~4(d)], suggesting that the longitudinal mode may be an 
elementary excitation of the ordered system in this regime.

\begin{figure}[t] 
\begin{center}
\includegraphics[width=0.34\textwidth]{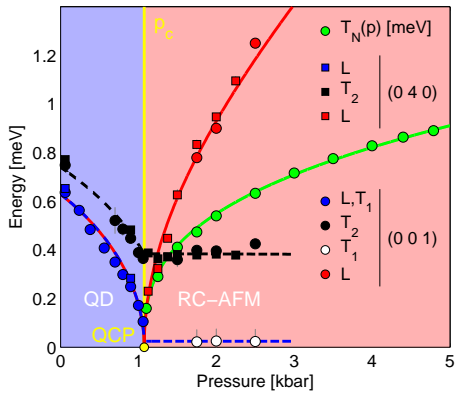}
\end{center}
\caption{Summary of INS results for the gaps of all three triplet 
excitations as functions of pressure at $T = 1.85$ K. Data for $T_{\rm N} (p)$
from Ref.~\cite{Rueegg04}. Modes L and T$_1$ 
are degenerate within experimental resolution at $p < p_c$. Red symbols 
show the longitudinal mode L at $p > p_c$. Solid and dashed lines are 
theoretical fits.}
\label{fig3}  
\end{figure} 

The fits of excitation energies and intensities in Figs.~3 and 4 are 
obtained from a theoretical model based on the bond--operator technique 
\cite{Matsumoto04}. A specific representation for the weakly ordered 
phase is provided by the superposition of singlet and triplet states 
on each dimer bond: the QD state is described by the singlet component 
($|s \rangle$), and the properties of the RC phase by an additional
triplet component. For the pressure--induced QPT, the wave function 
$|\sigma_i \rangle$ of a dimer may be written as  
\begin{equation}
|\sigma_i \rangle = \left[ \cos \theta |s \rangle + \sin 
\theta e^{i {{\mbox{\boldmath$Q_{\rm AF}$\unboldmath}} \cdot 
{\mbox{\boldmath$r_i$\unboldmath}}}} |t_z \rangle \right]
\end{equation}
where $\theta$ increases monotonically with pressure from 
0 at $p = p_c$ to $\pi/4$ for perfect antiferromagnetism, 
${\mbox{\boldmath$Q_{\rm AF}$\unboldmath}}$ is the ordering wave vector,
and ${\mbox{\boldmath$r_i$\unboldmath}}$ the position of dimer $i$. 
The triplet mixing coefficient $\sin \theta$ is the sole parameter 
determining all the physical properties of the ordered state ($T_{\rm N} 
(p)$, $m_{s} (p) = g \mu_{\rm B} \sin \theta / \sqrt{2}$, $\Delta_{\rm L} 
(p)$), and is specified entirely by the pressure evolution of the 
superexchange parameters. The emergence of the longitudinal mode is 
contained naturally in this theoretical framework. 

\begin{figure}[t] 
\begin{center}
\includegraphics[width=0.45\textwidth]{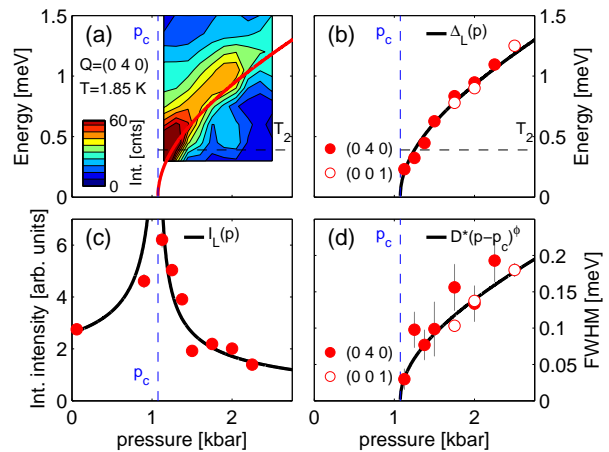}
\end{center}
\caption{\small Longitudinal mode in the pressure--controlled RC 
phase. (a) INS intensity as a function of energy for predominantly 
longitudinal fluctuations (red peaks, Fig.~2) measured at 
${\mbox{\boldmath$Q$\unboldmath}} = $(0 4 0). (b) Longitudinal 
mode gap $\Delta_{\rm L} (p)$: the black curve obtained from the
theoretical description has a square--root form, $\Delta_{\rm L} (p) 
\propto (p - p_c)^{1/2}$. (c) Integrated scattering intensity, which 
is inversely proportional to the gap for $p > p_c$. (d) Full width at 
half maximum height: here the black line is a guide to the eye, with 
fitted exponent $\phi = 0.5 \pm 0.1$. }
\label{fig4}  
\end{figure} 

The problem of modeling hydrostatic pressure effects in TlCuCl$_3$ is 
underconstrained. We have fitted the data by assuming both an increase 
of $J_2$ (an interdimer coupling in the $a$--$c$ plane \cite{Matsumoto04}) 
and a reduction of $J$. Either change in isolation acts to close the gap 
and to alter the dispersion, making this linear at the band minimum at 
the QPT, where a perfectly SU(2)--symmetric system would have three spin 
waves. The evolution of the mode gaps at $p < p_c$, and the ordered moment 
and longitudinal mode gap at $p > p_c$, are reproduced with the functional 
forms $J(p) = J + A_0 p + B_0 p^2$, $J_2(p) = J_2 + A_2 p + B_2 p^2$. 
The exponents of the transition are dictated by the linear terms, which 
were taken as $A_2 = - A_0 = 0.00660 {\rm kbar}^{-1}$, while the 
quadratic coefficients $B_2 = - B_0 = 0.00109 {\rm kbar}^{-2}$ were 
also necessary to ensure an adequate fit.

Similarly, the anisotropic interactions required to account for
the experimental observations may reside on the dimer bonds, on the 
interdimer bonds, or on both. In a minimal model where only $J$ is 
anisotropic, one may define uni-- and biaxial anisotropy parameters
$J_{xx}$ and $J_{yz}$ by $J_x = J + J_{xx}$, $J_{y,z} = J \pm J_{yz}$.
The conclusions obtained using interdimer exchange anisotropy are
qualitatively identical. The excitation gaps are very sensitive to 
this anisotropy, which can thus be deduced with extremely high precision 
from the INS data. The low--pressure data show two resolved mode energies, 
the best fit giving gaps $\Delta = 0.65$ meV and 0.79 meV. The separation 
of the upper mode (T$_2$) is reproduced by an easy--plane, uniaxial 
anisotropy $J_{xx} = 0.008 J$ for pure intradimer anisotropy 
($J_{xx}^{\prime} = - 0.004 J'$ for pure interdimer anisotropy). 
At $p > p_c$ we observe (a) one massive "spin--wave" (transverse, 
T$_2$) mode with gap $\Delta = 0.38$ meV, (b) one nearly massless 
transverse mode (T$_1$) with fitted gap 0.023 meV, and (c) one 
excitation which becomes higher--lying away from $p_c$ (L). (a) The 
gap of T$_2$ is in good agreement with the value 0.8\% ($-$0.4\%) for 
the uniaxial anisotropy component deduced at $p < p_c$. (b) The data 
correspond to a biaxial anisotropy of 0.002\% (-0.001\%), a value 
impossible to resolve at $p < p_c$, and are more appropriately 
considered as setting an effective upper limit on the possible mass 
of T$_1$. (c) The longitudinal mode shows a characteristic pressure 
evolution where the gap scales with the ordered moment and N\'eel 
temperature, following precisely the parameter--free fit in Fig.~4(b). 
The data is consistent with the same anisotropy for all pressures, and 
its value with that deduced from electron spin resonance measurements 
\cite{rkgto}.

The field--induced QPT, because it involves a U(1)--symmetric order 
parameter and quadratically dispersing bosons, has been described as 
a Bose--Einstein condensation (BEC) of the single magnon mode which 
becomes massless \cite{Giamarchi08}. Even for precisely uniaxial 
exchange anisotropy, the pressure--induced transition, where the 
magnon dispersion is linear, cannot qualify as a BEC and has the 
scaling exponents of a different universality class. However, from 
the structure of the theoretical description, the ordered phase 
remains a ''condensate of magnons''. In a further (related) contrast, 
at the pressure--induced QPT the Goldstone modes are explicitly those 
triplets not mixing with the singlet, while the linear singlet--triplet 
combination orthogonal to the ground state [Eq.~(2)] becomes massive.

Evidence for longitudinal excitations in quantum magnets has been 
reported in structures including the $S = 1$ chain compound CsNiCl$_3$ 
\cite{Kenzelmann02} and the $S = 1/2$ chain systems KCuF$_3$ \cite{rltn} 
and BaCu$_2$Si$_2$O$_7$ \cite{rzkmun}. However, the nature of the 
excitation spectra in these quasi--one--dimensional materials, 
which are dominated by a spinon--like continuum, is ambiguous to the 
extent that the existence of the longitudinal mode has been called into 
question \cite{rzkmun} on the grounds that it may decay into spin waves. 
Our continuous control of the ordered state allows a fully systematic
approach to $p_c$, and it is clear that the longitudinal mode shows 
no sign of a divergent decay at the QPT [Fig.~4(d)]. In fact the 
fitted exponent of the decay as a function of pressure, $\phi = 0.5
\pm 0.1$, matches closely the exponent of the gap. For strict equality, 
the mode could not be called truly elementary, but would be best described 
as ''critically well--defined''. That the mode should have precisely 
this nature at the QPT in a three--dimensional system was deduced 
in Ref.~\cite{raw} from the fact that $3 + 1$ is the upper critical 
dimension in this case. Further measurements over a range of temperatures 
are required to verify whether this result is a genuine example of quantum 
critical dynamics.

Our results have a direct connection to the properties of many other
quantum spin systems with strong fluctuation phenomena and partially
or entirely suppressed magnetism \cite{Sachdev08}. Excitation spectra 
have been measured in cases including the $S$ = 1 ("Haldane") chain 
and two--leg $S$ = 1/2 ladder (both QD), and the $S$ = 1/2 square 
lattice (RC, \cite{rrea}). The QD state of most interest in quantum 
magnetism is one where the singlets are no longer localized, the 
resonating valence--bond (RVB) state. While positional resonance may 
be the mechanism for suppression of order in some highly isotropic 
models, low--lying triplet states are a generic feature of any system 
near a magnetic QPT. Excitation spectra in the RVB 
framework have been discussed explicitly for the square lattice 
\cite{rhmoa}, and the same approach to other models would reveal 
emergent low--lying modes in RVB states. This type of investigation 
may be realized not only in spin systems under pressure, for example 
in the highly frustrated kagome geometry \cite{rhea}, but also in cold 
fermionic systems on optical lattices \cite{rtstz}. Finally, 
several theories for high--temperature superconductivity are based 
on the existence of a QPT \cite{Sachdev08,rs} occurring as a function 
of hole doping. Our results deliver the clear message that a complete 
account of the excitation spectrum is an integral part of establishing 
the validity of such a scenario. 

In summary, we have performed high--resolution neutron spectroscopy 
on the quantum antiferromagnet TlCuCl$_3$ over a range of applied 
pressure values across the magnetic quantum phase transition. We 
demonstrate the continuous evolution of the spin dynamics and drive 
the emergence of a critically well--defined longitudinal excitation 
(amplitude mode) of the ordered moment in the renormalized classical 
phase. A theoretical framework is developed which describes the measured 
excitations in all regions of the phase diagram. We use this systematic 
experimental control of the quantum state to illustrate in every detail 
the profound connection between the excitations of phases separated by 
a quantum critical point.

We thank T. M. Rice and M. Sigrist for invaluable contributions.
We are grateful to G. Aeppli, C. Batista, M. Vojta, and M. Zhitomirsky
for helpful comments. This work was performed in part at the Swiss
spallation neutron source, SINQ, at the Paul Scherrer Institute, and
was supported by the Swiss National Science Foundation, the NCCR MaNEP,
and the Wolfson Foundation.

\end{document}